# Broken symmetry and symmetry classification of magnetic ordering in doped crystals

B.R. Gadjiev

*International University of Nature, Society and Mans "Dubna",
141980, 19 Universitetskaya str., Dubna, Moscow Region, Russia*
gadjiev@uni-dubna.ru

**Abstract**
The problem of the spontaneous breaking of the symmetry in the crystals with variable composition is discussed. The character of the variation of the inhomogeneous superstructure depending on the substitutional atom concentration has been considered. The genesis of the structure and magnetic ordering in the compounds of the $U(Pd_xFe_{1-x})_2Ge_2$ type and perovskite of the $(Ln_yLn^*_{1-y})_{1-x}A_xMnO_3$ type is discussed.

### Introduction

The paper is devoted to the problem of the spontaneous broken symmetry in doped compounds.

Neutron diffraction research of crystals with variable composition has revealed the basic feature of the phase diagram of the doped compounds [1]. Thus with variation of substitutional atom concentration and external fields in doped structures the sets of modification of the structure and magnetic ordering are discovered [2, 3, 4, 5].

In the doped structures with an incommensurate phase with variations of substitutional atom concentration, the space group of symmetry of the structure being invariant, the variation of the wave vector of modulation of the structure is observed. As a rule, the wave vector of magnetic structure varies along the certain direction in a Brillouin zone, corresponding to the space group symmetry of the structure and the value variation of a wave vector occurs along the line, the end points have of which have the maximum symmetry of the structure. For example, the wave vector of modulation in the compounds $UPd_2Si_2$, $UPd_2Ge_2$ and $U(Pd_{0.98}Fe_{0.02})_2Ge_2$ are $\vec{q} = (0,0,0.662 \pm 0.010)c$, $\vec{q} = (0,0,1)c$ and $\vec{q} = (0,0,0.748 \pm 0.010)c$, respectively [2, 3]. Besides, with the increase of substitutional atom concentration the temperature width of the existence of an incommensurate phase is narrowed. In this context further we shall discuss the problems of magnetic ordering of the ThCr2Si2 type compounds [2].

We shall consider manganites of the $(Ln_yLn^*_{1-y})_{1-x}A_xMnO_3$ type (where Ln is a rare – earth and A is a alkaline – earth or a alkali metal), with a structure perovskite as an example of the structure without incommensurate phase. In these compounds depending on the concentration of substitutional atom the paramagnetic phase has either *Pnma* or $R\bar{3}c$ space group of symmetry [4, 5]. At low temperatures in perovskite like compounds one can observe ferromagnetic phase that undergoes a concentration phase transition to an antiferromagnetic phase with the increase of substitutional atom concentration. In the compound $(La_{0.25}Pr_{0.75})_{0.7}Ca_{0.3}MnO_3$ with an isotope $O^{16}$ with the increase of temperature the phase transition from ferromagnetic phase to antiferromagnetic one with wave vectors of the magnetic ordering $\vec{k} = (\frac{1}{2}00)$ and $\vec{k} = (\frac{1}{2}0\frac{1}{2})$ is observed. In the compounds $(La_{0.25}Pr_{0.75})_{0.7}Ca_{0.3}MnO_3$ with an isotope $O^{18}$ with the decrease of temperature the antiferromagnetic phase with wave vectors of ordering $\vec{k} = (\frac{1}{2}00)$ and $\vec{k} = (\frac{1}{2}0\frac{1}{2})$ remains invariable up to low temperature. Thus, the results of the neutron scattering researches of magnetic ordering in the perovskite type structures show that with variation of substitutional atom concentration, the space group symmetry in these compounds being invariant, we have a threshold variation of the wave vector of the magnetic ordering.

**The problem**

Knowledge of symmetry of the order parameter is necessary for the solution of the problem of spontaneous breaking of symmetry. Knowledge of space group symmetry of paramagnetic phase and wave vector of modulation of superstructure allows to define representation of the group of symmetry of the structure, according to which the order parameter will be transformed [6, 7].

The thermodynamic potential functional is constructed by the invariants of representation $D$ of the space group of high symmetry phase $G_0$, which determines the lowering of symmetry $G_0 \Rightarrow G_1$, where $G_1$ is a subgroup of group $G_0$. In the ideal crystal the equilibrium positions of the atoms are defined by the vector $\vec{R} = n_1\vec{a}_1 + n_2\vec{a}_2 + n_3\vec{a}_3 + \vec{r}_s$, where $n_1\vec{a}_1 + n_2\vec{a}_2 + n_3\vec{a}_3$ runs on the sites of the basic lattice, and $\vec{r}_s$ ($s = 1,…,S$) sets basis in an elementary cell, and the structure has one of 230 space groups. For definition of the order parameter symmetry in doped structures it is necessary to notice, that, strictly speaking, in contrast to the ideal crystals distributions of atoms have no space group of symmetry. Namely, the substitution of the atoms in the structure breaks the periodicity of the ideal crystals.

Let us discuss the influence of doping on the magnetic orderings in periodic structures, assuming that the doping, induces variation with accuracy up to on the atomic structure does not change the space group symmetry of the structure. The deformations in continuous medium caused by doping, could be describe by the vector of displacement and, hence, by the real symmetric tensor of deformation [7].

First we shall consider the case when the atoms of basis in an elementary cell are substituted whereas the atoms of the structures in lattice sites are not substituted. Substitution of atoms of the basis in various elementary cells in the structure doped in this way results in the variation of atom coordinates and thus weak deformation of the structure.

However when the atoms in the basic lattice sites are substituted one should expect strong distortion in the structure as the sites of a crystal lattice have a higher symmetry in the site as compared to the general position in an elementary cell.

Therefore for the definition of the order parameter symmetry in doped structures it is necessary to notice that, actually, the symmetry of doped structures is lower, than the symmetry of the similar structures, constructed by the identical spheres. However in theoretical consideration it is possible to introduce a geometrical characteristic — the order parameter, which describes this or that concrete lowering of the symmetry from the close-packed degenerated structure to the really observable derivative structure. While discussing theoretically the spontaneous symmetry breaking in doped crystals we should have it in mind that the approximation preservation of the space group of the structure, requires at least an account of the interaction of major order parameter with a secondary one, namely with deformation. Besides, the account of the energy, contributed by the substitution of atoms of the structure, is necessary [8].

Thus, irrespective of the way of the atom substitution of the structure the approximated preservation of space group of symmetry of doped crystals requires an account of the elastic energy $f_{elastic}$ in the thermodynamic potential functional $f[\eta, u, m]$ and the energy $mE_{impurity}[\eta]$ contributed by the substitutional atom of the structure. Thus the thermodynamic potential functional of doped structure is represented as

$$f[\eta, u, m] = f_{ideal}[\eta] + f_{elastic}[u] + mE_{impurity}[\eta] + f_{interaction}[\eta, u].$$

Here η is the order parameter, $u$ is the tensor deformation and $m$ is concentration of the doped atoms. The structure of inhomogeneous, characteristic of the derivative of the real structure, could be obtained from the analysis of the solution, determining the transition from the degenerated close-packed structure to the observable one.

**Field effects on inhomogeneous superstructures**

Let us consider the influence of a field on inhomogeneous superstructures. We could expect, that under the influence of a field the wave vector of superstructure would deviate from the initial symmetric direction.

In discussion of the incommensurate phase in ideal crystals on the basis of the Landau theory two cases are usually distinguished. In the first of them we have the symmetry of two (or more) component order parameter, which is supposed by the Lifshitz gradient invariant. In the second case it is absent for one component order parameter. These two cases are called incommensurate phases of type II and I respectively [9].

Let us define the character of the wave vector of the superstructure variation under the influence of a weak field in the case of type I transitions. In this case the problem is reduced to minimization of the thermodynamic potential functional, constructed according to symmetry arguments. Expansion of the thermodynamic potential contained the terms with interaction of order parameter with the deformation tensor

$$F = \frac{\alpha + ma}{2}\eta^2 + \frac{\beta + mb}{4}\eta^4 + (\sigma + cm)k\eta^2 + \frac{\kappa + me}{2}k^2\eta^2 + \frac{c}{2}\varepsilon_{zz}^2 + \delta\eta^2\varepsilon_{zz} + \gamma k\eta^2\varepsilon_{zz} \qquad (1)$$

where we take into account interaction $z$ – component of deformation tensor with order parameter only. The account others the component of deformation tensor, results additional renormalization of constant interaction δ and γ, which characterize interactions of order parameter with deformation. Here $\alpha = \alpha_0(T - T_0)$ is reduced temperature, $T_0$ is critical temperature. α, β, σ, k, and a, b, c, e are parameter of the thermodynamic potential functional corresponding to expansion $f_{ideal}[\eta]$ and $E_{impurity}[\eta]$, respectively. $c$ is the elastic module of the parent phase. From the equilibrium condition we obtain the values of the wave vector $\bar{k}$, the spontaneous deformation $\varepsilon_{zz}$ and the order parameter $\eta_0$.

$$k = \frac{-1}{\kappa + me}(\sigma + cm + \gamma\varepsilon_{zz}) \qquad (2)$$

$$\varepsilon_{zz} = \frac{1}{c}(\delta + \gamma k)\eta^2 \qquad (3)$$

$$\eta_0^2 = \frac{\alpha_0(T - T_c)}{\beta_0} \qquad (4)$$

Here $T_c = T_0 - \frac{ma}{\alpha_0} + \frac{1}{\alpha_0}(\frac{(\sigma + cm)^2}{\kappa + me} - \frac{\delta^2}{c})$ is the shift of the critical temperature and $\beta_0 = \beta + mb + \frac{2(\sigma + cm)\gamma\delta}{(\kappa + me)c}$.

Hence, the wave vector of modulation is a linear function of the spontaneous deformation, and with the variation $\varepsilon_{zz}$ the value of the wave vector varies. Besides, it is possible to show, that the critical temperature $T_c$ linearly depends on the components of the spontaneous deformation $T_c = \frac{1}{\alpha_0}[\frac{(\sigma + cm)^2}{\kappa + me} + \delta(2 - \frac{\gamma}{\kappa})\varepsilon_{zz}]$. Therefore it is possible to assert, that the wave vector of the superstructure in the doped structures is a linear function of a spontaneous deformation and of a substitutional atom concentration.

Let us consider the case of the phase transition of type II when the phase transition in the system is described by one component order parameter and therefore is transformed according to one-dimensional irreducible representation of the space group high symmetry phase. According to the structure 230-space groups the one-dimensional irreducible representation could correspond either to the center, or to the surface points of a Brillouin zone [10]. Suppose that the group $G_{\bar{k}} \subset G$ is a little group of the vector $\bar{k}$ and n is the number of rays in the star of the vector $\bar{k}$, and p is dimension of the irreducible representations of the little group $G_{\bar{k}}$. Then the dimension of irreducible representations $D_0(\bar{k})$ of the group $G$ is defined as $s = n \otimes p$. Hence, the one-dimensional irreducible representations of space group could arise only, when the little group of the vector $\bar{k}$ coincides with the group of symmetry of the structure. It is necessary to notice that the one-dimensional irreducible representations satisfy Lifshitz condition.

Let us define the character of the variation of a superstructure wave vector under the influence of the weak field, when the little group of the wave vector $\bar{k}$ coincides with group of symmetry of the structure. In this case Lifshitz invariant is absent, and the density of thermodynamic potential functional of the system is represented by the expression.

$$f = \frac{\alpha + am}{2}\eta^2 + \frac{\beta_0 + mb}{4}\eta^4 + \frac{\delta_0 + md}{2}(\frac{d\eta}{dx})^2 + \frac{\lambda_0 + lm}{2}(\frac{d^2\eta}{dx^2})^2 + \frac{\chi_0 + ms}{2}\eta^2(\frac{d\eta}{dx})^2 \quad (5)$$

Here $\alpha = \alpha_0(T - T_0)$ is reduced temperature, $T_0$ is critical temperature. $\beta_0$, $\delta_0$, $\lambda_0$, $\chi_0$ and $a$, $b$, $d$, $l$, $s$ are parameter of the thermodynamic potential functional corresponding to expansion $f_{ideal}[\eta]$ and $E_{impurity}[\eta]$, respectively. We introduce designation $\beta = \beta_0 + mb, \delta = \delta_0 + md, \lambda = \lambda_0 + lm$, $\chi = \chi_0 + ms$ and after minimization of $f$ the solutions for the order parameter are presented in the form:

(i)      if $\chi > 0$ and $\beta g_1 < 0$, $\eta = \rho snpx$ and the wave vector is $p^2 = -\frac{\beta g_1}{\chi(1+k^2)}$,

(ii)      if $\chi < 0$ and $\beta g_1 < 0$, $\eta = \rho dnpx$ and the wave vector is $p^2 = \frac{\beta g_1}{\chi(2-k^2)}$, (6)

(iii)      if $\chi < 0$ and $\beta g_1 > 0$, $\eta = \rho cnpx$ and the wave vector is $p^2 = -\frac{\beta \bar{g}_1}{\chi}$,

where $g_1$ is some combination of the thermodynamic potential parameter and $\bar{g}_1 = g_1/(2-k^2)$, and $k^2 \in [0,1]$. The account of the influence of the deformation results in variation of the thermodynamic potential parameters $\beta_0 = \beta - \frac{2f_0^2}{c}$, $\delta_0 = \delta + \frac{2e\kappa}{c}$ and $\chi_0 = \chi - \frac{\delta_0 f_0}{c}$, and also in the shift of the critical temperature $T_c^* = T_c - \frac{am}{\alpha_0} - \frac{2ef_0}{\alpha_0 c}$ [6].

The variation of the thermodynamic potential parameters could break the conditions, limiting realization of phases with the given modulation. Therefore there exist critical value deformations or concentration of doped atoms, that could result in threshold, variation of a wave vector modulation of superstructure.

### Magnetic orderings in compounds of the U(Pd$_{1-x}$Fe$_x$)$_2$Ge$_2$ type

The phases of symmetry $I4/mmm$ and $P4/mmm$ corresponding to real tetragonal structures could arise as a result of the phase transition from the latent phase with symmetry space group $Im3m$. The transition from the latent phase to the corresponding superstructure is described by an order parameter that is transformed according to the irreducible representation of the group *Im3m*, belonging to the point $\vec{k}_8 = (0,0,\frac{2\pi}{na})$ [6, 11].

The star of the vector $\vec{k}_8$ contains six rays, and the corresponding order parameter is transformed according to a six-dimensional irreducible representation $\tau_1(\vec{k}_8)$, induced by a totally symmetric small representation of the group of the wave vector $G_{\vec{k}_8} = C_{4v}$. The phases of the symmetry *I4/mmm* and *P4/mmm* arise for odd and even values $n \geq 3$, when only two rays of $\vec{k}_8$ are active. We stress that the vector $\vec{k}_8$ characterizes the line Δ, which begins in the Brillouin zone center Γ and ends in the vertex H. Thus the Γ and H points have the highest structural symmetry *Im3m*, and on the line Δ the symmetry is lowered to $C_{4v}$. Since the line Δ is internal to the Brillouin zone, the six-dimensional irreducible representation $\tau_1(\vec{k}_8)$ is passive and consequently the phase transition *Im3m* ⇒ *I4/mmm* in the general case is carried out through an incommensurate phase.

Let us consider the structure with a space group *I4/mmm* in the paramagnetic phase. In the Brillouin zone the line characterized by the vector $\vec{k}_{10} = (0,0,\frac{2\mu\pi}{\tau_z})$, begins in the center $\vec{k}_{14} = (0,0,0)$ and ends in the point $\vec{k}_{15} = (0,0,\frac{\pi}{\tau_z})$. *I4/mmm* is the little groups of the wave vectors $\vec{k}_{14}$ и $\vec{k}_{15}$. $C_{4v}$ is the little group of the vector $\vec{k}_{10}$. Let us consider two-dimensional irreducible representation $D^{*\vec{k}_{10}}$ of the space group *I4/mmm*, corresponding to the vector $\vec{k}_{10}$. Since the vector $\vec{k}_{10}$ is internal to Brillouin zone the irreducible representation $D^{*\vec{k}_{10}}$ does not satisfy Lifshitz condition and therefore the thermodynamic potential contains the Lifshitz invariant.

In accordance with the condition of invariance the thermodynamic potential functional is expressed as:

$$F = \frac{1}{d}\int_0^d f(x)dx, \quad f(x) = \frac{\alpha}{2}\rho^2 + \frac{\beta}{4}\rho^4 + \gamma\rho^n \cos n\varphi - \delta\rho^2(\frac{d\varphi}{dx}) + \frac{k}{2}\rho^2(\frac{d\varphi}{dx})^2, \quad (7)$$

where the parameters of the thermodynamic potential are the linear functions of concentration of doped atoms. With the increase of a doping concentration the temperature of the phase transition from the paramagnetic phase to the modulated one is lowered $T_i(m) = T_i(m=0) - am$.

Spatial dependence the phase of the order parameter is defined by Jacobi elliptic function

$$\varphi(x) = \frac{2}{n}am(px,\kappa), \quad (8)$$

where $\kappa \in [0,1]$. The temperature interval in which the incommensurate phase exists depends on the doping concentration according to the equation

$$T_i(m) - T_c = (\frac{k}{\gamma})^{\frac{2}{n-2}}(\frac{\beta}{\alpha_0})^{\frac{1}{2}}(\frac{\pi\delta}{4k})^{\frac{4}{n-2}} - am. \quad (9)$$

It is clear that as the doping concentration increases, the temperature region in which the incommensurate phase exists becomes narrower.

In the commensurate phase the expression for the magnetic moment in the *n*-th unit cell of the crystal in terms of the magnetic moment in the first cell is

$$\vec{M}_{\vec{n}} = \vec{M}_1 e^{i\vec{k}\vec{n}} + \vec{M}_1 e^{-i\vec{k}\vec{n}},$$

where $\vec{n}$ is the vector of the lattice, $\vec{k}$ is the wave vector of the superstructure. It follows from this expression that in the case $\vec{M}_1 = \frac{M}{2}(\vec{m}_1 + ip\vec{m}_2)$, $\vec{m}_1\vec{m}_2 = 0$ and $\vec{m}_1^2 = \vec{m}_2^2$ with the ellipticity parameter $p \neq 0$ the magnetic moment $\vec{M}_{\vec{n}} = M[\vec{m}_1 \cos\vec{k}\vec{n} - p\vec{m}_2 \sin\vec{k}\vec{n}]$, describes the elliptical helix in the space. If we set $p = 0$ we have a structure of the spin wave type $\vec{M}_{\vec{n}} = M\vec{m}_1 \cos\vec{k}\vec{n}$. For $\vec{m}_1 \parallel \vec{k}$ the last expression describes the *LSW* structure and for $\vec{m}_1 \perp \vec{k}$ it describes the *TSW* structure.

The one-dimensional irreducible representation of the group *I4/mmm* that corresponds to the vector $\vec{k}_{15} = (0,0,\frac{\pi}{\tau_z})$ determines a thermodynamic potential without Lifshitz invariant. This functional describes a structure of the *LSW* type. The equilibrium condition of the functional determines the antiferromagnetic ordering of the type η = ρ*snpx* and η = ρ*cnpx* and the phase with ferromagnetic ordering of the type η = ρ*dnpx*. The commensurate phase that corresponds to these phases is antiferromagnetically ordered and the magnetic cell is doubled.

### Magnetic orderings in compounds of the $(La_{0.25}Pr_{0.75})_{0.7}Ca_{0.3}MnO_3$ type

The phase of symmetry $O_h^1$ corresponding to $ABO_3$ type crystals could arise as a result of phase transitions from a latent phase with the symmetry space group $O_h^9$. The phase transition $O_h^9 \Rightarrow O_h^1$ is described by the order parameter that is transformed according to the irreducible representation of the group $O_h^9$ belonging to the point $\vec{k}_{12}$. In the framework of the Landau theory the real structures of perovskites $ABO_3$ could be obtained in the result of the phase transition described by the reducible representation $\tau_5(\vec{k}_{11}) \oplus \tau_8(\vec{k}_{13})$ of the space group $O_h^1$. Thus the symmetry analysis shows, that the phase with the group of symmetry *Pnma* = $D_{2h}^{16}$ corresponds to the solution of $(\eta_1, 0, 0, 0, \zeta_2, \zeta_3)$ kind, and the phase of symmetry $R\bar{3}c = D_{3d}^6$ corresponds to the solution of $(0, 0, 0, \zeta_1, \zeta_2, \zeta_3)$ kind, where η and ζ — the order parameter which is transformed according to the reducible representation $\tau_5(\vec{k}_{11}) \oplus \tau_8(\vec{k}_{13})$ of the group $O_h^1$ [7].

The powder X-ray diffraction patterns for all the samples $Pr_{0.7}Sr_{0.3-xx}MnO_3$ and $Pr_{0.7-xx}Sr_{0.3}MnO_3$ could be indexed with a rhombohedral perovskite structure and $R\bar{3}c$ space group for $x \leq 0.2$ in strontium deficient samples and for $x \leq 0.1$ for praseodymium deficient one. For the other value of $x$ the samples could be indexed in the orthorhombic structure with *Pbnm* space group. Magnetization measurements *versus* temperature shows that all the samples exhibit a magnetic transition when the temperature decreases [4].

Thus in the case of group $R\bar{3}c$ the thermodynamic potential functional is formed by the invariants of the reducible representation $A_{1g} \oplus A_{2g}$. If group *Pbnm* being a group symmetry of the structure, the thermodynamic potential functional consists of the invariants of the reducible representation $A_g \oplus B_{1g}$. The integer basis of the invariants of the representations $A_{1g} \oplus A_{2g}$ and $A_g \oplus B_{1g}$ coincide. Therefore, in both cases the symmetry consideration leads to a thermodynamic potential functional of the form

$$f(x) = \frac{\alpha_2}{2}\eta^2 + \frac{\beta_2}{4}\eta^4 + \frac{a}{2}u^2 + \frac{b}{4}u^4 + \gamma\eta^2 u \tag{10}$$

The equation of state corresponding to (10)

$$\eta(\alpha_2 + \beta_2\eta^2 + 2\gamma u) = 0$$

$$u(a + bu^2) + \gamma\eta^2 = 0 \tag{11}$$

have a solution

(I) $\quad u = 0, \eta = 0$

(II) $\quad u^2 = -\dfrac{a}{b}, \eta = 0$ (12)

(III) $\quad u \neq 0, \eta \neq 0$.

In phases (I) and (II) the average values of magnetization are equal to zero. In phase (III) it is not equal to zero and consequently the phase is ferromagnetic.

Let us consider the structure with the space group $D_{2h}^{16}$ in paramagnetic phase. In this case $D_{2h}^{16}$ is the little group of the vectors $\vec{k}_{20}, \vec{k}_{24}, \vec{k}_{19}$ [10]. For the symmetry classification of the magnetic ordering in the doped structure $(La_{0.25}Pr_{0.75})_{0.7}Ca_{0.3}MnO_3$ it is necessary to consider the reducible representation $D^{\vec{k}_{20}} \oplus D^{\vec{k}_{22}} \oplus D^{\vec{k}_{19}} \oplus A_g$, which according to symmetry arguments leads to the model of the thermodynamic potential functional as

$$f(x) = \frac{\alpha}{2}(\varphi_1^2 + \varphi_2^2) + \frac{\beta}{4}(\varphi_1^2 + \varphi_2^2)^2 + \frac{\upsilon}{2}\varphi_1^2\varphi_2^2 + \frac{g_3}{2}\psi^2(\varphi_1^2 + \varphi_2^2) + \frac{\alpha}{2}\psi^2 + \frac{\beta_1}{4}\psi^4 + \\ \frac{\alpha_2}{2}\eta^2 + \frac{\beta_2}{4}\eta^4 + \frac{a}{2}u^2 + \frac{b}{4}u^4 + \gamma\eta^2 u + \zeta(\varphi_1^2 + \varphi_2^2)u + \xi\psi^2 u$$ (13)

Here $\alpha = \alpha_0(T - T_N)$ and $\alpha_2 = \alpha_{20}(T - T_c)$, where $T_N$ and $T_c$ are the temperatures of the transition to antiferromagnetic and ferromagnetic phase, accordingly. The minimization of the thermodynamic potential according to the fields $\varphi_1$, $\varphi_2$, $\eta$, $\psi$ and $u$ results in the system of the equations

$$\varphi_1[\alpha + \beta(\varphi_1^2 + \varphi_2^2) + \upsilon\varphi_2^2 + g_3\psi^2 + 2\zeta u] = 0$$

$$\varphi_2[\alpha + \beta(\varphi_1^2 + \varphi_2^2) + \upsilon\varphi_1^2 + g_3\psi^2 + 2\zeta u] = 0$$

$$\psi[\alpha + \beta_1\psi^2 + g_3(\varphi_1^2 + \varphi_2^2) + 2\xi u] = 0$$ (14)

$$\eta[\alpha_2 + \beta_2\eta^2 + 2\gamma u] = 0$$

$$au + bu^3 + \gamma\eta^2 + \zeta(\varphi_1^2 + \varphi_2^2) + \xi\psi^2 = 0$$

Thus for the initial paramagnetic phase we obtain (I) $\varphi_1 = \varphi_2 = \psi = \eta = u = 0$. The phase corresponding to the solution, (II) $\varphi_1 = \varphi_2 = \psi = \eta = 0$, $u \neq 0$ is also paramagnetic.

The solution of the system of the equations (14), corresponding to the antiferromagnetically- ordering phase with the wave vectors of modulation $\vec{k}_{20}$ and $\vec{k}_{22}$, has the form (III) $\varphi_1 \neq 0$, $\varphi_2 \neq 0$, $\psi \neq 0$, $u \neq 0$, $\eta = 0$.

The solution of the system of the equations (14), corresponding to the ferromagnetically ordering phase has the form (IV) $\varphi_1 = 0, \varphi_2 = 0, \psi = 0, u \neq 0, \eta \neq 0$.

It is easy to show, that the value of displacement in phases (III) and (IV) are not equal. As the spatial variation of the deformation tensor defines the deformation potential therefore nonlinear electron-phonon interaction is the reason of the destruction of the magnetic ordering in these phases

In $(La_{0.25}Pr_{0.75})_{0.7}Ca_{0.3}MnO_3$ with an isotope $O^{18}$ with the lowering of temperature the change of phases is observed (I) $\Rightarrow$ (III), whereas in the same compounds with complete substitution $O^{16} \Rightarrow O^{18}$ with lowering of temperature the change of phase (I) $\Rightarrow$ (III) $\Rightarrow$ (IV) is observed.

**Conclusion**

The wave vector of the magnetic ordering of the structure deviates from the initial symmetrical position under the influence of the field. This variation continuously occurs along the definite direction in Brillouin zone for incommensurate superstructures with a wave vector of general position. For the wave vectors, having the maximum symmetry of the structure, the variation of the wave vector of the superstructure has a threshold character. The tetragonal structure with the space group $I4/mmm$, which is the symmetry group of paramagnetic phase of compounds $ThCr_2Si_2$ type, arises as a result of a structural phase transition from the close–packed latent phase with space group zone of group $I4/mmm$. The phase transition $Im3m \Rightarrow I4/mmm$ is carried out through an intermediate incommensurate phase. The transformation property of the order parameter in the doped structures of the $ThCr_2Si_2$ type is determined by the irreducible representation corresponding to the line $\Omega$, which is characterized by $Im3m$. With the increase of substitutional atom concentration, the variation of the value of vector $\vec{k}$ occurs on a line, which begins in the center $\Gamma$ and ends in the surface point H in a Brillouin zone of group $I4/mmm$. Vector $\vec{k}$ being doped, variation of its value takes place on the line $\Omega$.

For the functional with the Lifshitz invariant the solution of the *LSW* type arises in the case when the ellipticity parameter is equal to zero and the magnetic moments are directed along the modulation wave vector. The variation of character of the magnetic ordering occurs in the points $\vec{k}_{14} = (0,0,0)$ and $\vec{k}_{14} = (0,0,\frac{\pi}{\tau_z})$. The appropriate functional does not contain the Lifshitz invariant, and the magnetic orderings are characterized by the solutions $\eta = \rho snpx$, $\eta = \rho dnpx$ and $\eta = \rho cnpx$.

The observable structure perovskite with the chemical formulae $ABO_3$ could be obtained in the result of phase transition described by the reducible representation $\tau_5(\vec{k}_{11}) \oplus \tau_8(\vec{k}_{13})$ of the group symmetry $O_h^1$. For example, with variation of substitutional atom concentration either phase with group of symmetry $Pnma = D_{2h}^{16}$, or phase with group of symmetry $R\bar{3}c = D_{3d}^6$ arise. The threshold variation of the wave vector of the magnetic ordering is characteristic for the perovskite with chemical formulae $ABO_3$. The structures of perovskites $Pr_{0.7}Sr_{0.3-x}{}_xMnO_3$ and $Pr_{0.7-x}{}_xSr_{0.3}MnO_3$ under observation are the function of the substitutional atom concentration and are either $R\bar{3}c = D_{3d}^6$, or $Pnma = D_{2h}^{16}$. In both structures the magnetic ordering is described by the reducible representation. At low temperatures these compounds have a ferromagnetic ordering phase, that with the increase of temperature is transformed into a paramagnetic phase. The observed structures in more complicated doped compounds $(La_{0.25}Pr_{0.75})_{0.7}Ca_{0.3}MnO_3$ with on isotope $O^{18}$ and isotope $O^{16}$ have a space group symmetry $Pnma$. In the compounds $(La_{0.25}Pr_{0.75})_{0.7}Ca_{0.3}MnO_3$ with an isotope $O^{18}$ at low temperatures one can observe ferromagnetically ordered phase that with the increase of temperature is transformed into an antiferromagnetic phase, then to paramagnetic one. The compounds $(La_{0.25}Pr_{0.75})_{0.7}Ca_{0.3}MnO_3$ with an isotope $O^{16}$ at low temperatures has an antiferromagnetic phase that with the increase of temperature is transformed into a paramagnetic phase.